\documentclass[%
 reprint,
 amsmath,amssymb,
 aps,prc,
floatfix,
]{revtex4-2}

\usepackage{graphicx}
\usepackage{dcolumn}
\usepackage{bm}
\usepackage{amsmath}
\usepackage{nccmath}
\usepackage[dvipdfmx]{hyperref}
\usepackage[dvipdfmx]{xcolor}
\hypersetup{colorlinks=true,linkcolor=blue,citecolor=blue,urlcolor=blue}
\usepackage{mathtools}
\usepackage{natbib}

\begin{document}


\title{Modes of massive nucleon transfer appearing in quasifission processes \\for collisions of superheavy nuclei}

\author{S. Amano$^{1}$, Y. Aritomo$^{1}$ and M. Ohta$^{2}$}
\affiliation{$^{1}$Kindai University, Higashi-Osaka, Osaka 577-8502, Japan}
\affiliation{$^{2}$Konan University, Kobe, Hyogo 658-8501, Japan}

\date{\today}

\begin{abstract}
\noindent{\bf Background:} It is challenging to distinguish between fusion-fission  and quasifission experimentally. To determine the characteristics of quasifission processes associated with dominant phenomena in heavy-ion collisions is important for estimating precisely the fusion cross section, which is relevant to the synthesis of new elements. We classified fusion-fission and quasifission processes theoretically in the past for an accurate assessment of the fusion cross section \lbrack\href{https://doi.org/10.1016/j.nuclphysa.2004.08.009}{\color{blue}Nucl.\;Phys.\;A\;{\bf 744},\;3\;(2004)}\rbrack. 
However, no detailed analysis focused on each process was performed. \\
{\bf Purpose:} In this work, we aimed to analyze the dynamical characteristics of quasifission processes in terms of the Langevin equation model. 
We specify the quasifission processes, and analyze the scission configuration. Finally, we clarify the origin of several modes included in quasifission.\\
{\bf Method:} The calculation framework is the multidimensional dynamical model of nucleus-nucleus collisions based
on the Langevin equations.\\
{\bf Results:} It is shown that several quasifission modes exist leading to different fragment deformations. The time scale of the quasifission process differs for several different modes. Each scission configuration and total kinetic energy also differ.\\
{\bf Conclusions:} The different quasifission modes are caused by the neck relaxation controlling the mass drift toward symmetry. This means that it is possible to discuss the time-dependent functional form of the neck parameter $\epsilon$ for the quasifission process in the framework of the dynamical model based on Langevin equations.
\end{abstract}

\maketitle


\section{Introduction}
Heavy nuclear collisions associated with the large mass transfer between reaction partners have been attracting considerable attention in the last several decades as incomplete fusions and quasifission (QF) processes \cite{PhysRevC.36.115, PhysRevC.53.1734, ITKIS2015204, PhysRevC.90.054608, PhysRevC.94.054613, PhysRevC.99.014616}. Various phenomena can be seen in these reactions, depending on the induced angular momentum, i.e., complete fusion producing compound nucleus (CN), QF, deep inelastic collision (DIC) and quasielastic collision (QEC). 
The theoretical identification of these phenomena has been traditionally carried out on the basis of the geometrical difference in the impact parameter (or corresponding the angular momentum) of the incident channel. At low impact parameters, complete fusions are dominant. Many QF events are included in the reactions at low-medium impact parameters. For medium-high impact parameters, the dominant reaction type is DIC. When impact parameters is high, QEC is observed. 
In the quantum-based dinuclear system model \cite{PhysRevC.64.024604}, QF is described on the basis of the property of the diffusion of nucleons in the mass asymmetry of reaction partners. Multinucleon transfer in the DIC process has been studied by the Langevin-type approach \cite{PhysRevC.96.024618}.
Here, we apply the Langevin-type approach to investigate the dynamical characteristics of the QF process.

QF processes are considered to be a bridge between DIC, where much relative kinetic energy is transferred to their thermal energy, and CN formation leading to evaporation residue (ER) products after possible particle emissions \cite{ITKIS2015204}.  The estimation of the ER cross section in the superheavy mass region has been paid much attention in the challenge of new element synthesis. In this sense, the study of the QF mechanisms, that is, the dynamical characteristics of QF and the separation of QF and CNF (CN fission), is important for the precise estimation of the ER cross section. 
In particular, the study of the dynamics of QF process can be expected to open a new field for the reaction mechanism in the superheavy mass region because the process is related to the mass transfer and the shell structure. 

To date, the characteristics of QF have been discussed in the paper on the total kinetic energy (TKE) of fission fragments including the long-standing problem of energy transfer to the kinetic energy of fragments and in the investigation of the time scale of the mass drift to the mass symmetric direction \cite{PhysRevC.36.115}.

In this paper, the dynamical characteristics of the QF process are presented by the analysis of dynamical Langevin trajectories using the unified Langevin equation model \cite{zagrebaev2007potential}. This Langevin-type approach was developed by Zagrebaev $\it{et~ al.} $ \cite{zagrebaev2007potential} for the analysis of heavy-ion collision from the entrance stage to the fused system including fission processes. We have revealed the existence of QF modes in the reactions of the $^{48}$Ca projectile and the actinide target system. These QF modes are investigated in connection with their time scale, the distribution of the fragment deformation and with the time evolution of neck formation. In terms of the trajectory analysis of the QF process, we show the reason for the appearance of the mode. 

In the following section, a brief introduction of the Langevin-type approach is presented. 
The dynamical QF characteristics of the $^{48}$Ca + $^{249}$Bk system at $E$$_\text{c.m.}$=213.05 MeV are described in Sect. 3, where the detailed analyses of the properties of the QF mode and the reason for their appearance are presented. Our concluding remarks are given in the final section.

\section{Framework}
\subsection{Potential energy surface}
We adopt the dynamical model that is similar to the unified model \cite{Zagrebaev_2007}. First, the initial stage of the nucleon transfer reactions consists of two parts: 
(1) the system is calculated in the ground state of the projectile and target because the reaction is too fast for the nucleons to reconfigure a single-particle state. (2) Then, the former system relaxes to the ground state of the entire composite system, which changes the potential energy surface to an adiabatic one. 
Therefore, we consider the time evolution of the potential energy from the diabatic one $V_\text{diab}\left(q\right)$ to the adiabatic one $V_\text{adiab}\left(q\right)$. 
Here, $q$ denotes a set of collective coordinates representing a nuclear shape. The diabatic potential is calculated by a folding procedure using effective nucleon-nucleon interaction \cite{Zagrebaev_2005,Zagrebaev_2007,zagrebaev2007potential}. 
However, the adiabatic potential energy of the system is calculated using an extended two-center shell model \cite{zagrebaev2007potential}. 
Then, we connect the diabatic and adiabatic potential energies with a time-dependent weighting function $f\left(t\right)$ as follows:
%
\begin{eqnarray}
&&V(q,t)=V_\text{diab}\left(q\right)f\left(t\right)+V_\text{adiab}\left(q\right)\left[1-f\left(t\right)\right], \\
&&f\left(t\right)=\exp{\left(-\frac{t}{\tau}\right)}.
\label{pot}
\end{eqnarray}
%
Here, $t$ is the interaction time and $\tau$ is the relaxation time in the transition from the diabatic potential energy to the adiabatic one. 
We use the relaxation time $\tau=10^{-22}~\text{s}$ proposed in Refs. \cite{bertsch1978collision,CASSING1983467,PhysRevC.69.021603}. 

We use the two-center parameterizations as coordinates to represent nuclear deformation \cite{maruhn1972asymmetrie,sato1978microscopic}.
To solve the dynamical equation numerically and avoid the huge computation time, 
we strictly limited the number of degrees of freedom and employed three parameters as follows: 
$z_{0}$ (distance between the centers of two potentials), 
$\delta$ (deformation of fragment), and $\alpha$ (mass asymmetry of colliding nuclei);
$\alpha=\frac{A_{1}-A_{2}}{A_{1}+A_{2}}$, 
where $A_{1}$ and $A_{2}$ not only stand for the mass numbers of the target and projectile, respectively \cite{Zagrebaev_2005,ARITOMO20043}, but also are then used to indicate the mass numbers of the two fission fragments. $A_\text{CN}$ is the mass number of the compound nucleus.
As shown in Fig.~1 of Ref. \cite{maruhn1972asymmetrie}, the parameter $\delta$ is defined as $\delta=\frac{3\left(a-b\right)}{2a+b}$, where $a$ and $b$ represent the half  length of the ellipse axes in the $z_{0}$ and orthogonal to $z_{0}$ directions, respectively. 
We assume that each fragment has the same deformation in the first step. 
In addition, we use scaling to minimize the computation time and use the coordinate $z$ defined as $z=\frac{z_{0}}{R_\text{CN}B}$, where $R_\text{CN}$ denotes the radius of the spherical compound nucleus and the parameter $B$ is defined as $B=\frac{3+\delta}{3-2\delta}$.

The adiabatic potential energy is defined as
\begin{eqnarray}
V_\mathrm{{adiab}}\left(q,L,T\right) \qquad \qquad \qquad \qquad \qquad \qquad \qquad \qquad \nonumber \\
=V_\text{LDM}\left(q\right)+V_\text{SH}\left(q,T\right)+V_\text{rot}\left(q,L\right), \qquad  
\label{adipot}
\end{eqnarray}
where $V_\text{LDM}$ and $V_\text{SH}$ are the potential energy of the finite-range liquid drop model and the microscopic energy that takes into account the temperature dependence, respectively. The simplified symbol $V_\text{LDM}$ and the symbol $V_\text{SH}$ are described as  
\begin{eqnarray}
V_\text{LDM}\left(q\right)=E_\text{S}\left(q\right)+E_\text{C}\left(q\right), \qquad \\
V_\text{SH}\left(q,T\right)=E_\text{shell}^{0}\left(q\right)\Phi\left(T\right),\qquad \\
E_\text{shell}^{0}\left(q\right)=\Delta E_\text{shell}\left(q\right) + \Delta E_\text{pair}\left(q\right).\qquad
\end{eqnarray}
The symbols $E_\text{S}$ and $E_\text{C}$ stand for generalized surface energy \cite{PhysRevC.20.992} and Coulomb energy, respectively. 
The symbol $E_\text{shell}^{0}$ indicates the microscopic energy at $T$ = 0, which is calculated as the sum of the shell
correction energy $\Delta E_\text{shell}$ and the pairing correlation correction
energy $\Delta E_\text{pair}$. $T$ is the temperature of the compound nucleus calculated from the intrinsic energy of the composite system.
$\Delta E_\text{shell}$ is calculated by the Strutinsky method \cite{STRUTINSKY19681, RevModPhys.44.320}
from the single-particle levels of the two-center shell model
potential \cite{maruhn1972asymmetrie,suek74,10.1143/PTP.55.115} as the difference between the sum of
single-particle energies of occupied states and the averaged
quantity.
$\Delta E_\text{pair}$ is evaluated in the BCS approximation as described in Refs. \cite{RevModPhys.44.320, NILSSON19691}. The averaged part of the pairing correlation energy is calculated assuming that the density of single-particle
states is constant over the pairing window. The pairing strength constant is related to the average gap parameter $\tilde{\Delta}$ by solving the gap equation in the same approximation and adopting $\tilde{\Delta} = 12/ \sqrt{A}$ suggested in \cite{NILSSON19691} by considering the empirical results for the odd-even mass difference \cite{PhysRevC.90.054609}. 
The temperature dependence factor $\Phi\left(T\right)=\exp\left(-\frac{E^{\ast}}{E_\text{d}}\right)$ is explained in Ref. \cite{ARITOMO20043}, where $E^{\ast}$ indicates the excitation energy of the compound nucleus. $E^{\ast}$ is given as $E^{\ast}=aT^{2}$, where $a$ is the level density parameter. The shell damping energy $E_\text{d}$ is selected as 20 MeV. This value is given by Ignatyuk et al. \cite{ignatyuk1975phenomenological}. 
$V_\text{rot}$ is the centrifugal energy generated from the total angular momentum $L$. We obtain
\begin{eqnarray}
V_\text{rot}\left(q,L\right) \qquad \qquad \qquad \qquad \qquad \qquad \qquad \qquad \qquad \nonumber \\
=\frac{\hbar^{2}\ell\left(\ell+1\right)}{2\mathcal{I}\left(q\right)}+\frac{\hbar^{2}L_{1}(L_{1}+1)}{2\Im_{1}(q)}+\frac{\hbar^{2}L_{2}(L_{2}+1)}{2\Im_{2}(q)}. \quad 
\end{eqnarray}
Here, $\mathcal{I}\left(q\right)$ and $\ell$ represent the moment of inertia of the rigid body with deformation $q$ and the relative orientation of nuclei and relative angular momentum respectively. The moment of inertia and the angular momentum  for the heavy and light fragments are $\Im_{1,2}$ and $L_{1,2}$, respectively.

The neck parameter $\epsilon$ entering in the two-center parametrization was adjusted to reproduce the available data, assuming different values between the entrance and exit channels of the reactions \cite{YAMAJI1987487}. In our study, we use $\epsilon = 1 $ for the entrance channel and  $\epsilon = 0.35$ for the exit channel. We assume the following time dependence of Eq. (4), expressed in terms of the characteristic relaxation time of the neck $t_{0}$ and the variance $\Delta_{\epsilon}$: 
   
\begin{eqnarray}
V_\text{LDM}(q,t)&=&V_\text{LDM}(q,\epsilon=1) f_{\epsilon}(t) \nonumber \\
&+&V_\text{LDM}(q,\epsilon=0.35) [1- f_{\epsilon}(t)], \nonumber
\label{}
\end{eqnarray}
 
\begin{eqnarray}
&&f_{\epsilon}(t) = \frac{1}{1+\exp(\frac{t-t_{0}}{\Delta_{\epsilon}})}.
\label{}
\end{eqnarray}


\subsection{Dynamical equations}
We perform trajectory calculations of the time-dependent unified potential energy \cite{Zagrebaev_2005,Zagrebaev_2007,ARITOMO20043} using the multidimensional Langevin equation \cite{Zagrebaev_2005,ARITOMO20043,PhysRevC.80.064604} as follows : 
\begin{gather}
\frac{dq_{i}}{dt}=\left(m^{-1}\right)_{ij}p_{j}, \nonumber \\
\frac{dp_{i}}{dt}=-\frac{\partial V}{\partial q_{i}}-\frac{1}{2}\frac{\partial}{\partial q_{i}}\left(m^{-1}\right)_{jk}p_{j}p_{k}-\gamma_{ij}\left(m^{-1}\right)_{jk}p_{k} \nonumber \\
+g_{ij}R_{j}\left(t\right), \nonumber \\
\frac{d\vartheta}{dt}=\frac{\ell}{\mu_{R}R^{2}}, \nonumber \\
\frac{d\varphi_{1}}{dt}=\frac{L_{1}}{\Im_{1}}, \nonumber \\
\frac{d\varphi_{2}}{dt}=\frac{L_{2}}{\Im_{2}}, \nonumber \\
\frac{d\ell}{dt}=-\frac{\partial V}{\partial\theta}-\gamma_\text{tan}\left(\frac{\ell}{\mu_{R}R^{2}}-\frac{L_{1}}{\Im_{1}}a_{1}-\frac{L_{2}}{\Im_{2}}a_{2}\right)R
\nonumber \\
+Rg_\text{tan}R_\text{tan}\left(t\right), \nonumber \\
\frac{dL_{1}}{dt}=-\frac{\partial V}{\partial\varphi_{1}}+\gamma_\text{tan}\left(\frac{\ell}{\mu_{R}R^{2}}-\frac{L_{1}}{\Im_{1}}a_{1}-\frac{L_{2}}{\Im_{2}}a_{2}\right)a_{1}  \nonumber \\
-a_{1}g_\text{tan}R_\text{tan}\left(t\right), \nonumber \\
\frac{dL_{2}}{dt}=-\frac{\partial V}{\partial\varphi_{2}}+\gamma_\text{tan}\left(\frac{\ell}{\mu_{R}R^{2}}-\frac{L_{1}}{\Im_{1}}a_{1}-\frac{L_{2}}{\Im_{2}}a_{2}\right)a_{2}  
\nonumber \\
-a_{2}g_\text{tan}R_\text{tan}\left(t\right). 
\end{gather}
The collective coordinates $q_{i}$ represent $z, \delta$, and $\alpha,$ the symbol $p_{i}$ denotes  momentum conjugated to $q_{i}$, and $V$ is the multidimensional potential energy. The symbol $\vartheta$ indicates the relative orientation of nuclei. $\varphi_{1}$ and $\varphi_{2}$ stand for the rotation angles of the nuclei in the reaction plane, $a_{1,2}=\frac{R}{2}\pm\frac{R_{1}-R_{2}}{2}$ is the distance from the center of the fragment to the middle point between the nuclear surfaces, and $R_{1,2}$ is the nuclear radii. The symbol $R$ is the distance between the nuclear centers. 
The total angular momentum $L=\ell+L_{1}+L_{2}$ is preserved. The symbol $\mu_{R}$ is reduced mass, and $\gamma_\text{tan}$ is the tangential friction force of the colliding nuclei. 
Here, it is called sliding friction. 
The phenomenological nuclear friction forces for separated nuclei are expressed in terms of $\gamma_\text{tan}$ and $\gamma_{R}$ for sliding friction and radial friction using the Woods-Saxon radial form factor described in Refs. \cite{Zagrebaev_2005}. 
Sliding and radial friction are described as $\gamma_\text{tan}=\gamma_\text{t}^{0}F\left(\xi\right)$ and $\gamma_{R}=\gamma_{R}^{0}F\left(\xi\right)$, where the radial form factor $F\left(\xi\right)=\left(1+\exp^{\frac{\xi-\rho_{F}}{a_{F}}}\right)^{-1}$. $\gamma_\text{t}^{0}$ and $\gamma_{R}^{0}$ being the model parameters employ 0.1 $\times~10^{-22}$ MeV s fm$^{-2}$ and 100 $\times~10^{-22}$ MeV s fm$^{-2}$, respectively. $\rho_{F} \approx$ 2 fm and $a_{F} \approx$ 0.6 fm are also the model parameters determined in Ref. \cite{Zagrebaev_2005}, and $\xi$ is the distance between the nuclear surfaces $\xi=R-R_\text{contact}$, where $R_\text{contact}=R_{1}+R_{2}$ \cite{Zagrebaev_2005}. 
The phenomenological friction for the radial direction is switched to the one-body friction in the mononucleus stage.
$\gamma_{R}$ is described to consider the kinetic dissipation according to the surface friction model \cite{FROBRICH1998131}
The radial friction is calculated as $\gamma_{zz}=\gamma_{zz}^\text{{one}}+\theta\left(\xi\right)\gamma_{R}$. 
For the mononuclear system, the wall-and-window one-body dissipation $\gamma_{zz}^\text{{one}}$ is adopted for the friction tensor \cite{BLOCKI1978330,RAYFORDNIX1984161,RANDRUP1984105,Feldmeier_1987,CARJAN1986381,carj92,PhysRevLett.70.3538,20067}. 
$\theta\left(\xi\right)$ is smoothing function swiched the phenomenological friction to that of a mononuclear system as follows $\theta\left(\xi\right)=\left(1+\exp^{-\frac{\xi}{0.3}}\right)^{-1}$ \cite{Zagrebaev_2005}.
$m_{ij}$ and $\gamma_{ij}$ stand for the shape-dependent collective inertia and friction tensors, respectively. 
We adopted the hydrodynamical inertia tensor $m_{ij}$ in the Werner-Wheeler approximation for the velocity field \cite{PhysRevC.13.2385}. 
The one-body friction tensors $\gamma_{ij}$ are evaluated within the wall-and-window formula \cite{RANDRUP1984105, PhysRevC.21.982}.
The normalized random force $R_{i}\left(t\right)$ is assumed to be white noise: $\langle R_{i} (t) \rangle$ = 0 and $\langle R_{i} (t_{1})R_{j} (t_{2})\rangle = 2 \delta_{ij}\delta (t_{1}-t_{2})$. 
According to the Einstein relation, the strength of the random force $g_{ij}$ is given as $\gamma_{ij}T=\sum_{k}{g_{ij}g_{jk}}$.
We start trajectory calculations from a sufficiently long distance between both nuclei \cite{ARITOMO20043}. 
Thus, we use the master equation that takes into account the nucleon transfer for slightly separated nuclei \cite{Zagrebaev_2005,Zagrebaev_2005}. 
When the separated nuclei are achieved to the mononucleus state, the neck window of the contact nuclei is sufficiently opened; therefore, the nucleon transfer that occurs in the separated nuclei is almost neglected. Hence, the evolution of the mass asymmetry parameter $\alpha$ switches from the master equation to the Langevin equation 
in accordance with the procedure described in Ref. \cite{ARITOMO20043}.


\section{Results}
\subsection{QF trajectories depending on the initial orbital angular momentum}

\begin{figure*}[t]
\centering
\includegraphics[scale=0.45]{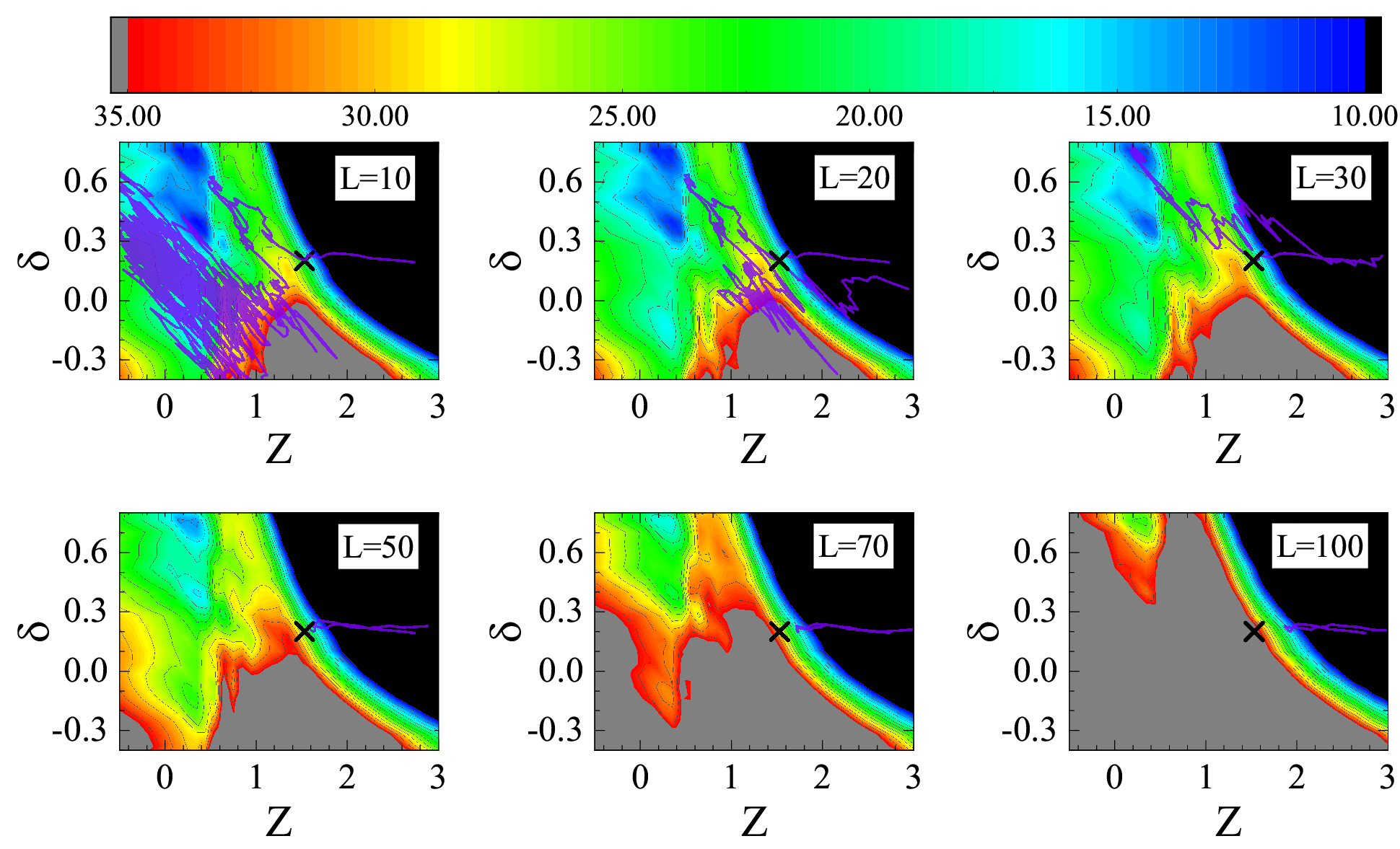}
\caption{The sample trajectories and the $z-\delta$ plane potential energy surface ($\alpha = 0.78$) for each orbital angular momentum ($L$) in the $^{26}$Mg + $^{230}$Th reaction. The calculation starting point is \{z, $\delta$\} = \{2.85, 0.20\}. The $\times$ marks denote the contact point. 
}\label{fig1}
\end{figure*}

To review the induced angular momentum dependence of the QF process, the sample trajectories projected on the $z-\delta$ plane of the potential energy surface for several  orbital angular momenta in the reaction $^{26}$Mg + $^{230}$Th ($\alpha=0.78$) at $E$$_\text{c.m.}$ = 125.00 MeV are shown in Fig. 1. The trajectory starts from \{z, $\delta$\} = \{2.85, 0.20\}. In the case of $L=10\hbar$, it can be seen that the trajectory invades the fusion area. 
The details of the fusion area are explained in Ref. \cite{ARITOMO20043}. 
The fusion area is defined as $\lbrace|\alpha|<0.3,\ \delta<-0.5z+0.5\rbrace$ in the present calculation. The trajectory for $L=10\hbar$ does not go towards the direction of fission ($+z$), because it is trapped in the ground state pocket even after ending the calculation. 
For the $L=20 \hbar$ case, however, the trajectory, after passing through the contact point marked by $\times$, does not enter the fusion area, and finally, the trajectory moves towards the direction of fission ($+z$). 
This behavior of the trajectory is categorized as the QF reaction. The trajectory for $L=30\hbar$ also shows the property of the QF process. At a higher value of $L$, the barrier indicated by the grey region begins to grow and it is difficult for the trajectories of yet higher $L$ to approach the contact point because of the increasing centrifugal potential energy. These behaviors of trajectories belong to QEC or DIC. In this paper, we treat the QF process as follows.
%
%
%
%
%
\subsection{Trajectory distribution on the deformation space}

To discuss the dynamics of the fusion-fission process, we introduce the probability distribution of the system in the deformation space.
To this end, we segment the coordinate space with $\Delta z =0.01, \Delta \delta =0.025,$ and 
$\Delta \alpha=0.001$. We define the distribution as an ensemble of Langevin trajectories.
We follow a trajectory as a function of time, and we increase the event number at each segment 
when the trajectory passes through that segment. By generating many trajectories, we 
construct a distribution of events on the deformation space \cite{PhysRevC.85.044614}.


Figure 2 shows the distribution of all trajectories in the $z-\delta$ plane for the reaction of $^{48}$Ca + $^{244}$Pu, $^{48}$Ca + $^{249}$Bk, and $^{74}$Ge + $^{208}$Pb systems with $E$$_\text{c.m.}$ = 206.91 MeV, $E$$_\text{c.m.}$ = 213.05 MeV, and $E$$_\text{c.m.}$ = 286.82 MeV, respectively. 
Here, the Langevin calculations are stopped at $t=10^{-19}$ s.
The $\delta$ distributions of the deformation of fragments at the scission point are also appended to the figure as histograms for $L=0\hbar, ~50\hbar$, and $100\hbar$. The trajectories start at the point marked with a star \{z, $\delta$\} = \{2.85, 0.20\} for the $^{48}$Ca + $^{244}$Pu and $^{48}$Ca + $^{249}$Bk systems, and \{z, $\delta$\} = \{2.85, 0.00\} for the $^{74}$Ge + $^{208}$Pb system. The initial value $\delta=0$ is set for the spherical nuclear case. The contact point is denoted as $\times$.

\begin{figure*}[t]
\centering
\includegraphics[scale=0.65]{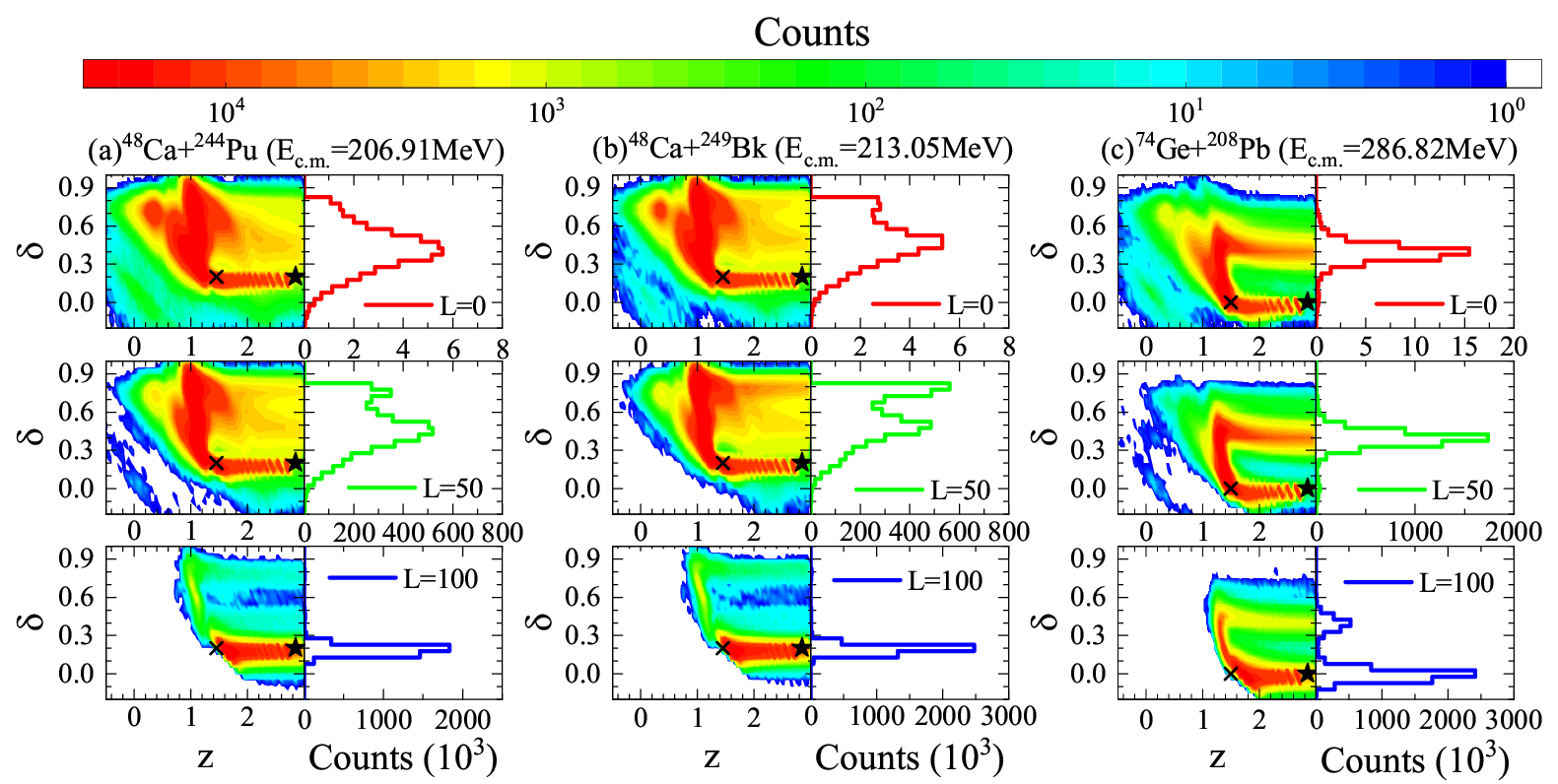}
\caption{The distribution of all trajectories and the $\delta$ distribution of the deformation of fragments at scission point for the induced orbital angular momentum $L=$0, 50, 100$\hbar$. The contact point and the starting point of the trajectories are marked by $\times$ and star, respectively. 
}\label{fig2}
\end{figure*}

\begin{figure}[t]
\flushleft
\includegraphics[scale=0.28]{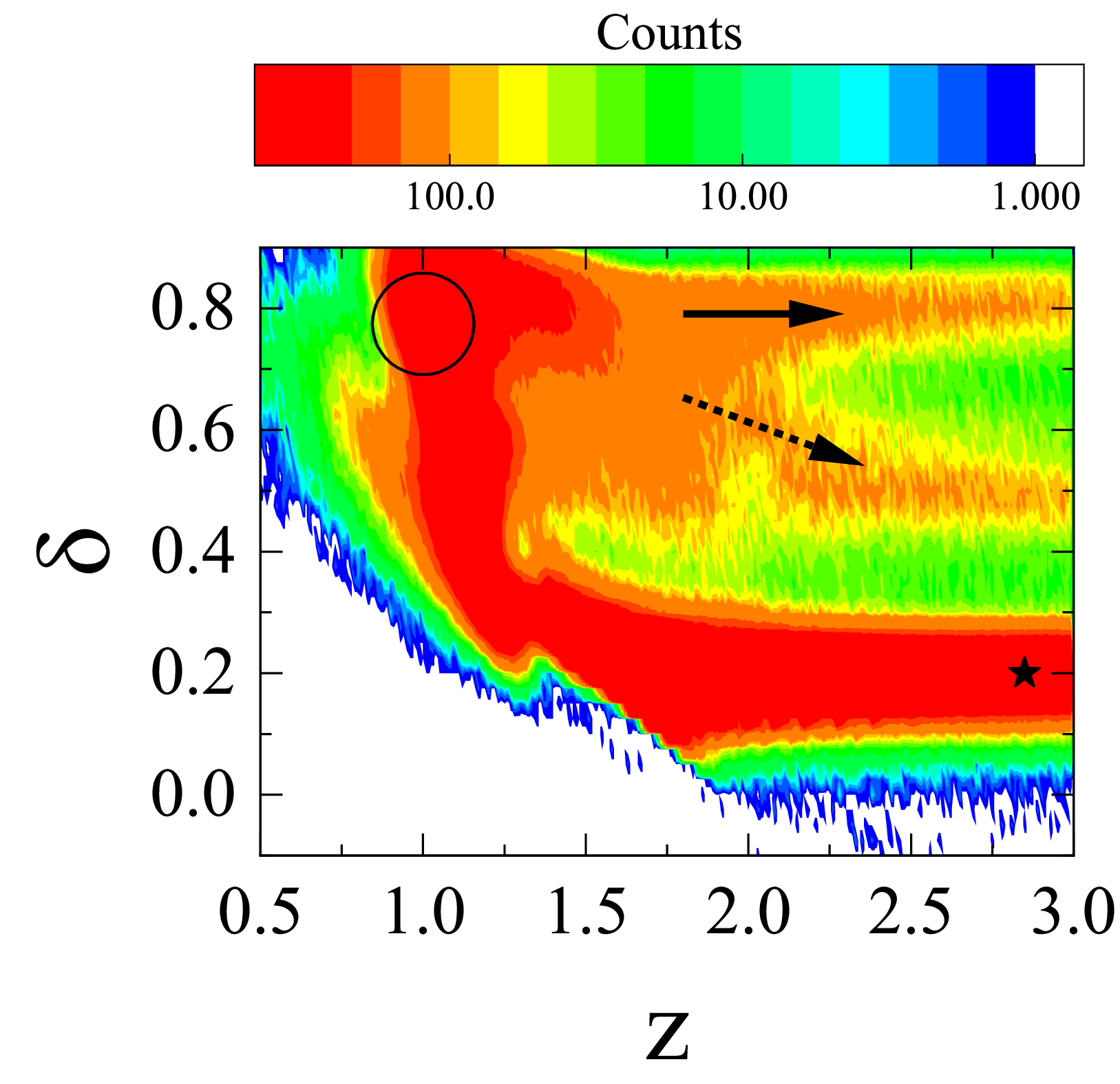}
\caption{The trajectory distribution for $^{48}$Ca + $^{249}$Bk system with $E$$_\text{c.m.}$ = 213.05 MeV and $L = 50\hbar$. Starting at star point \{z, $\delta$\} = \{2.85, 0.20\}, passing through the bifurcation point of paths marked with the symbol BP($\circ$), the trajectory evolves toward different area indicated by arrows. The path along the solid arrow is named ``path1'' and the dashed arrow ``path2''.
}\label{fig3}
\end{figure}
In Figs.~2(a) and 2(b), for the central collision ($L=0\hbar$) the trajectories near the scission point distribute around $\delta \approx 0.4$, but for $L=50\hbar$, the distribution has a structure with two peaks. The different quasifission modes appear in this angular momentum region. In the case of the $^{48}$Ca + $^{249}$Bk system, the structure of the $\delta$ distribution appears even for the low partial wave, and the bifurcation of the trajectory can be seen clearly. As the Coulomb interaction is stronger in this system, the approach to the fusion area is inhibited. The mode that appeared in the $\delta$ distribution has a different time scale, as discussed later.
In contrast, in the $^{74}$Ge + $^{208}$Pb system, no bifurcation in the quasifission path can be seen because of the entrance memory of the strong shell effect of the target $^{208}$Pb. From the starting point (×), the fragment deformation increases only up to $\delta \approx 0.6$ owing to the stiffness of  $^{208}$Pb. Therefore, the trajectory follows a similar area leading to the same fragment deformation around $\delta \approx 0.4$. 
We select a typical example of the QF path bifurcation from Fig.~2(b). Figure 3 shows an enlarged map of the trajectory distribution for the $^{48}$Ca + $^{249}$Bk system ($L=50\hbar$). From the starting point indicated by the star, the trajectory moves to the area of large $\delta$ up to 0.8 where the fragments are stretched in preparation of fission. Then, the fission path bifurcates at the point \{z, $\delta$\} = \{1.3, 0.8\}, and each path evolves along the arrows. We call the trajectory sustaining $\delta=0.8$ ``path1'' indicated by the solid arrow, and the other trajectory along the dashed arrow ``path2''.
The interesting point is the time scales of these two paths, which lead to the different fragment mass distributions, as shown in the next section.
\subsection{Two quasifission modes in massive nucleon transfer}
\begin{figure*}[t]
\centering
\includegraphics[scale=0.55]{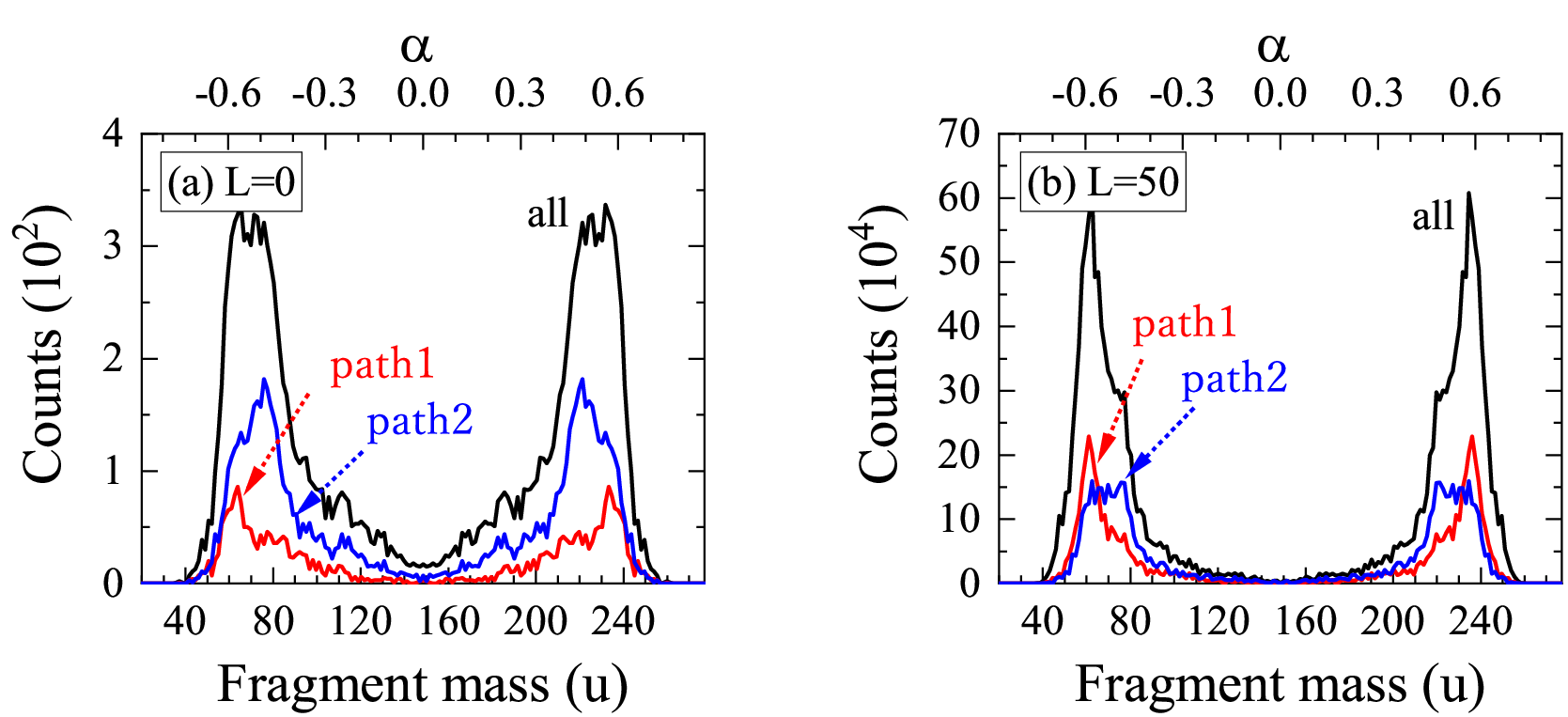}
\caption{The fission fragment mass distribution for $^{48}$Ca + $^{249}$Bk system with $E$$_\text{c.m.}$ = 213.05 MeV and $L=50\hbar$. The distributions in ``path1'' and ``path2'' are plotted separately by the red and the blue line, respectively. The panel (a) is for $L=0\hbar$ and (b) for $L=50\hbar$.
}\label{fig4}
\end{figure*}
\begin{figure*}[htbpt]
\centering
\includegraphics[scale=0.9]{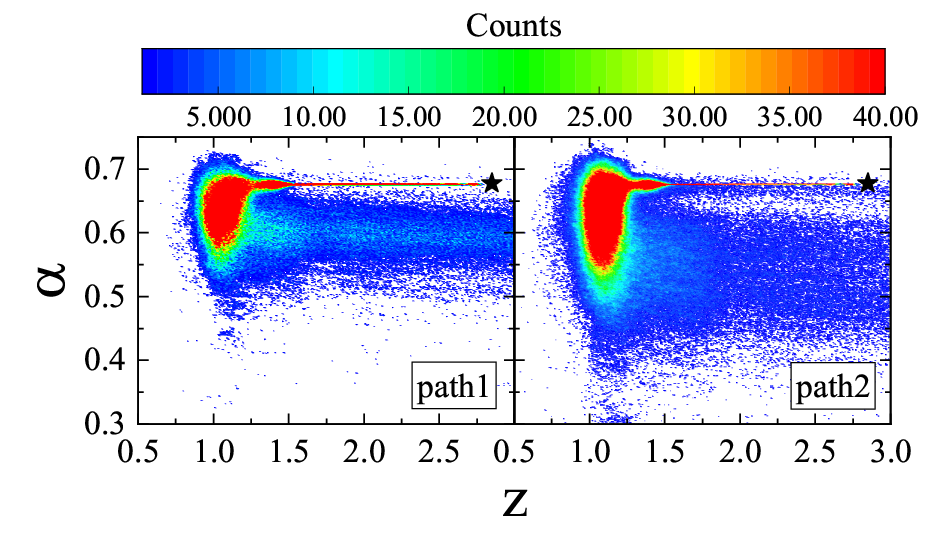}
\caption{The trajectory distribution of ``path1'' and ``path2'' drawn in $z-\alpha$ plane for $^{48}$Ca + $^{249}$Bk system with $E$$_\text{c.m.}$ = 213.05 MeV and $L=50\hbar$. The starting point of the trajectories are denoted by star symbol.
}\label{fig5}
\end{figure*}
\begin{figure}[htbp]
\flushleft
\includegraphics[scale=1.0]{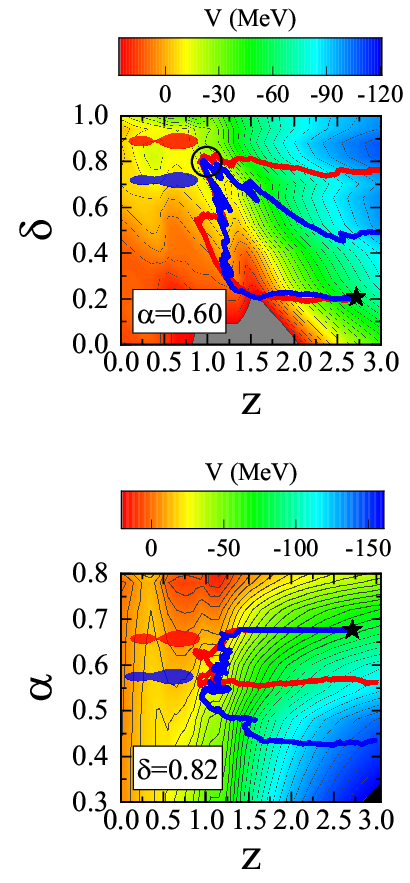}
\caption{The feature of the trajectory bifuraction and the shape of fissioning nuclei for $^{48}$Ca + $^{249}$Bk system with $E$$_\text{c.m.}$ = 213.05  MeV and $L=50\hbar$. The bifurcation point and the starting point of the trajectories are indicated by $\circ$ and the star, respectively. 
}\label{fig6}
\end{figure}
\begin{figure}[htbp]
\flushleft
\includegraphics[scale=0.53]{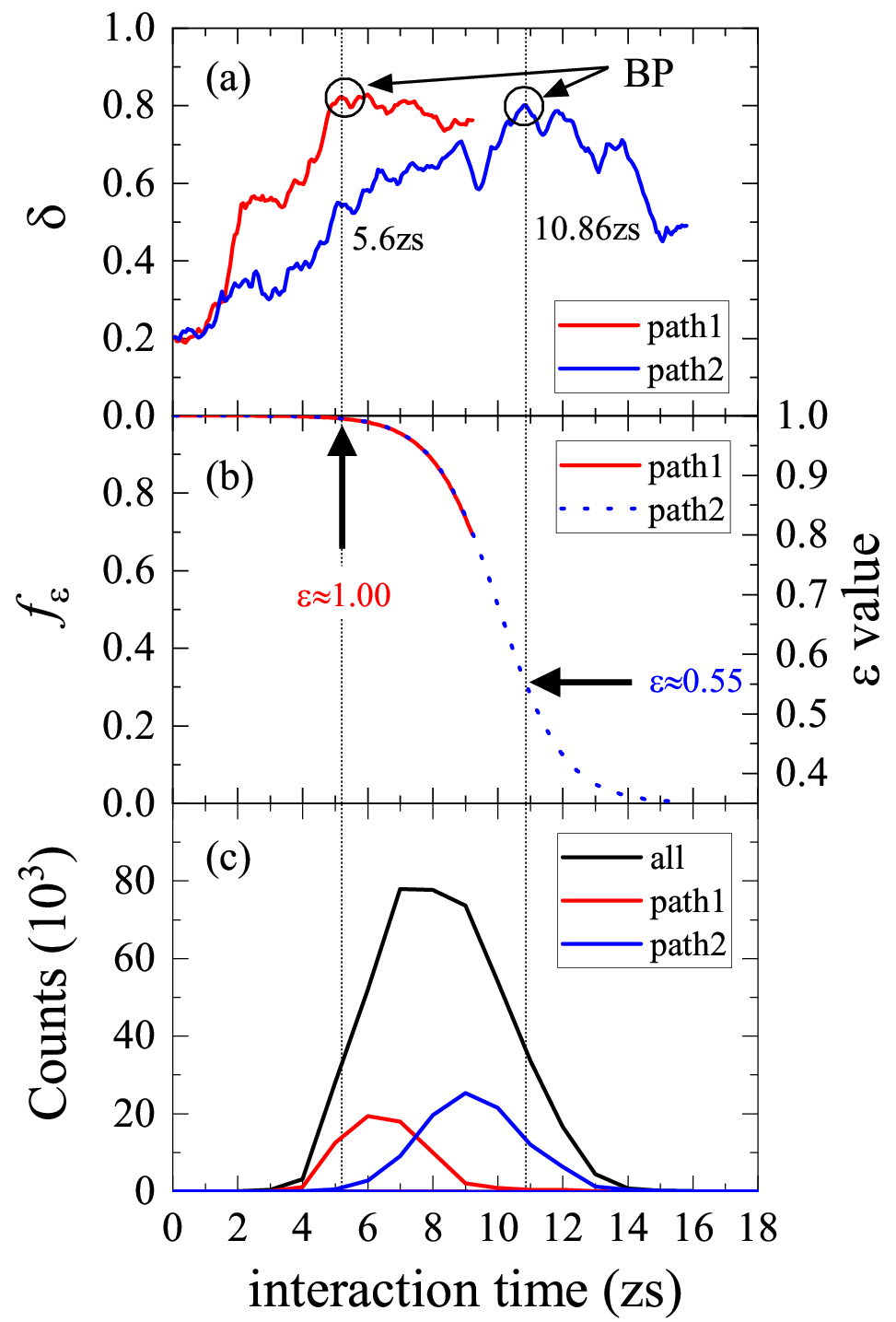}
\caption{The temporal evolution of the trajectory in panel (a), 
the neck parameter $\epsilon$ in (b) for the two different paths in the $^{48}$Ca + $^{249}$Bk system with $E$$_\text{c.m.}$ = 213.05 MeV and $L=50\hbar$. The distributions of the time duration reaching BP for different paths are plotted in (c). In all panels, vertical dotted lines indicate the position of the time reaching BP labeled in panel (a).
}\label{fig7}
\end{figure}
\begin{figure*}[t]
\centering
\includegraphics[scale=0.6]{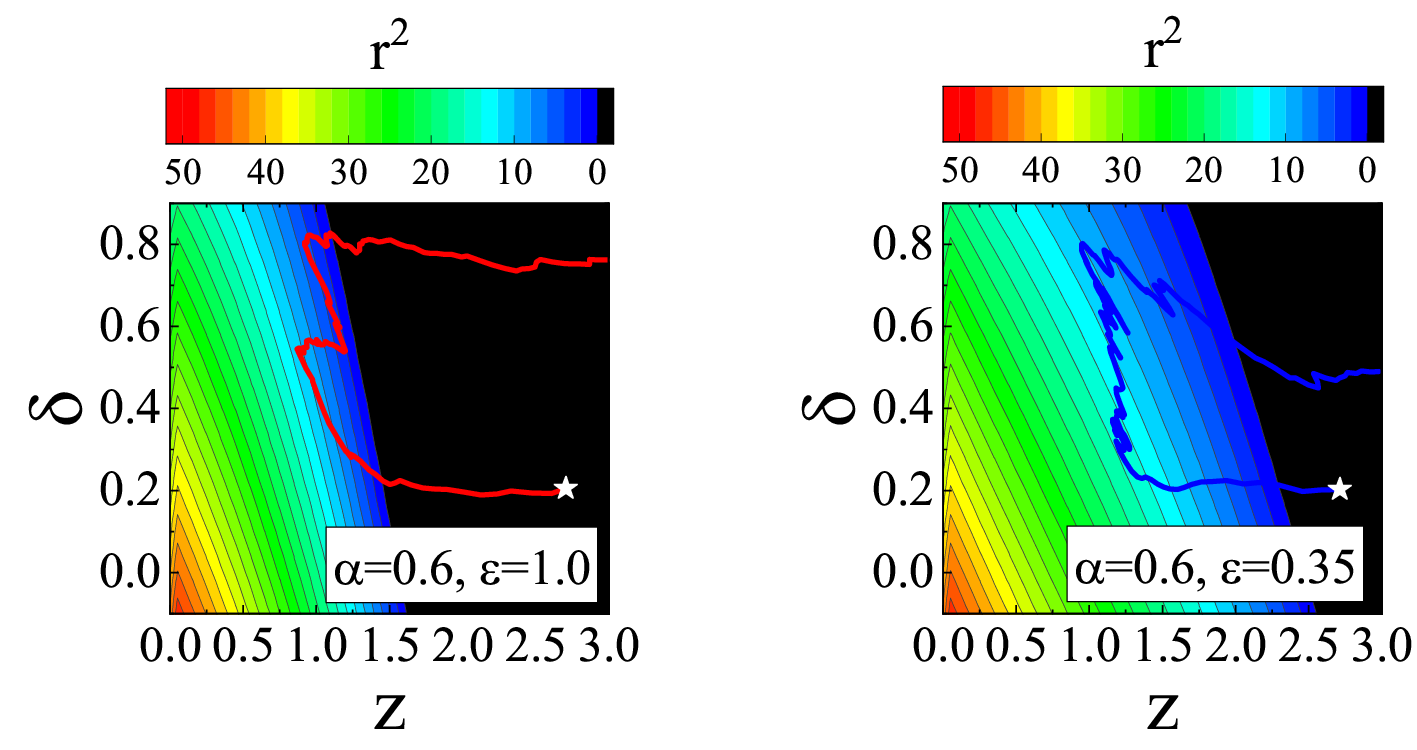}
\caption{The contour map of the square of neck radius $r^2$ in the $^{48}$Ca + $^{249}$Bk system same as Fig. 7. The trajectory of ``path1'' (left) and ``path2'' (right) are overwritten. The trajectories are started at the point marked by star. 
}\label{fig8}
\end{figure*}
\begin{figure}[t]
\flushleft
\includegraphics[scale=0.33]{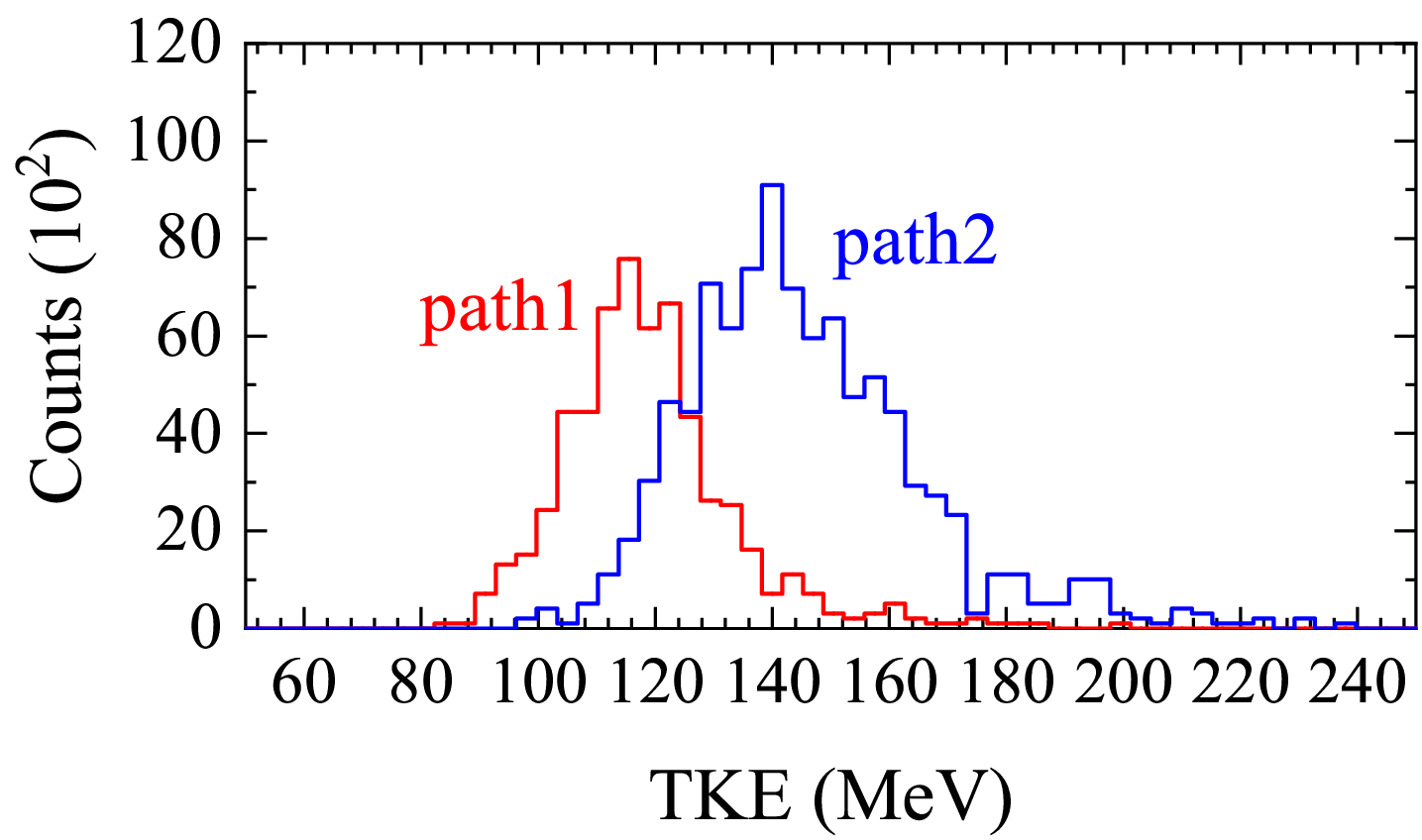}
\caption{The TKE distribution in the $^{48}$Ca + $^{249}$Bk system same as Fig. 7. The red and blue lines indicate ``path1'' and ``path2'', respectively. 
}\label{fig9}
\end{figure}
%
Figure 4 shows the fragment mass distributions for ``path1'' and ``path2''. The components of ``path1'' and ``path2'' correspond to the deformation ranges of $0.6<\delta<0.8$ and $0.4<\delta<0.6$ at the scission point, respectively. 
The main peak of the fragment distributions is located far outside the symmetric region of $A$$_\text{CN}$$/2 \pm20\text{u}$. Therefore, the normalized measure of the degree of mass drift toward the symmetry defined by Shen $\it et~ al.$ \cite{PhysRevC.36.115} $\Delta A/\Delta A_\text{max}$ is less than $\approx 0.3$, where $\Delta A/\Delta A_\text{max}=\frac{1}{2}(A_\text{t}-A_\text{p})$ and $\Delta A =A_\text{f}-A_\text{p}$. $A_\text{f},~A_\text{t}$, and $A_\text{p}$ are the fragment, target, and projectile mass, respectively.
Since both QF modes have a large elongation, no sign of the spherical fission fragment such as Pb can be seen.
However, the mass distribution shows a subtle difference. 
The peak of mass distribution for ``path2'' shifts to the mass symmetric region, in comparison with that of ``path1''. 
It is inferred that the mass drift is larger than that in ``path1'', as can be seen from the trajectory distribution plotted in the $z-\alpha$ plane shown in Fig.~5, where the trajectory distribution of ``path2'' extends to the lower value of $\alpha$.
We further investigate the dynamics around the bifurcation point (BP) of two modes near $\delta=0.8$. The feature of the trajectory bifurcation and the shape of fissioning nuclei are drawn in Fig.~6. The illustrations of fragment shape for ``path1'' and ``path2'' are distinguished by red and blue colors, respectively. The striking fact is in the role of neck radius for each path. The small neck radius of ``path1'' inhibits the mass drift at this point. On the one hand, the nuclear shape of ``path2'' has a large neck radius and the mass drift can be sustained after passing BP. As a result, the fragment mass distribution for ``path2'' shifts further to the mass symmetric region.
%
%
These differences originate from the fluctuation in the time scale of the reaction. 
As shown in Fig.~7(a), 
``path2'' belongs to the time-consuming trajectory to reach BP in comparison with ``path1''. The evolution degree of deformation in each path shows the sample results. The time that the sample trajectories of  ``path1' and ``path2'' reach BP are $\approx 5.6~\text{zs}$ and $\approx 10.8~\text{zs}$. 
The distribution of the time duration for each path to reach the area of BP is shown in Fig.~7(c). Owing to the random fluctuation of the trajectory combined in the potential energy surface (PES), the time duration to obtain BP also fluctuates. It takes $\approx 9~\text{zs}$ on average for ``path2''. In addition, the important point is in the behavior of the time evolution of the neck parameter $\epsilon$ in Fig.~7(b). The trajectory of ``path1'' reaches the scission point before the neck parameter decreases sufficiently toward the recommended value for fission $\epsilon \approx0.35$ \cite{YAMAJI1987487} in Eq. (8).
%
It is mentioned that the value can be extracted  from the analyses of the mass and energy distributions of the fission fragments \cite{zagrebaev2007potential}. In Fig.~7, we employ $t_{0}=9.0\times 10^{-21}$ s and $\Delta_{\epsilon}=1.0\times 10^{-21}$ s in Eq. (8).

%
%
It is clear that, as shown in Fig.~7(b), the neck parameter $\epsilon$ for each path differs at BP, $\it{i.e.}$,  $\epsilon \approx 1.0$ for ``path1'' and $\epsilon \approx 0.55$ for ``path2''. The neck radius near the scission point is strongly related with the time scale of the trajectory mentioned above. It is inferred that sufficient relaxation of the neck condition further accelerates the mass drift. For the trajectory taking a long time to reach BP, such as ``path2'', the value of $\epsilon$ is decreased to 0.55. In contrast, if the trajectory reaches BP in a short time, such as ``path1'', neck relaxation is incomplete and mass drift stops at that point.

This situation can be seen clearly in Fig.~8, where the contour map of the neck cross section is drawn in the $z-\delta$ plane overwriting the trajectory of each path. We can see the that neck cross section vanished at BP for ``path1'' but still exists for ``path2''. Therefore, in ``path2'', the mass drift continues after passing BP owing to the decrease in $\epsilon$, as shown in Fig.~7.

Here, we investigate the TKE of fission fragments to distinguish two modes. The scission configuration is defined as the shape with the neck radius equal to zero \cite{PhysRevC.88.044614}. The TKE is expressed as
\begin{eqnarray}
&\text{TKE}=V_\text{coul}+E_\text{pre},\qquad \\
&V_\text{coul}=e^2\frac{Z_{1}Z_{2}}{d_{sci}},\qquad \\
&E_\text{pre}=\frac{1}{2}\left(m^{-1}\right)_{ij}p_{i}p_{j},
\end{eqnarray}
where $E_\text{pre}$ and $V_\text{coul}$ denote the prescission kinetic energy and the prescission Coulomb energy, respectively. $e^{2}=1.44$ MeV fm, $Z_{1}, Z_{2}$ are the charge of each fragment, and $d_\text{sci}$ is the distance between the centers of mass of light and heavy parts of the nucleus at the scission point. The average of $E_\text{pre}$ over all fission events is equal to 7.03 MeV. Thus, the main contribution
to the TKE comes from the Coulomb repulsion of fission fragments.


%

Figure 9 shows the calculation results of  TKE shown by each path at $L=50\hbar$.
The TKE of each path differs owing to different $d_\text{sci}$ values. The average TKE values of each path are $\langle \text{TKE}_\text{path1}\rangle$ = 119 MeV and $\langle \text{TKE}_\text{path2}\rangle$ = 145 MeV. The lower TKE corresponds to a more elongated scission configuration (``path1''). The TKE values shown by ``path1'' and ``path2'' are different, and two modes clearly exist in QF.

\section{CONCLUSIONS}
The dynamical characteristics of the QF processes were investigated in terms of the Langevin equation model. We paid attention to the QF processes outside the fragment mass of $A$$_\text{CN}$$/2~\pm$ 20u and the inside DIC processes with several nucleon transfers. The $^{48}$Ca + $^{249}$Bk system was taken as an example. 
Detailed dynamical analysis of the QF trajectory in the Langevin calculation was performed. 
It was found that there exist two QF modes with different fragment mass and deformations. Results of the trajectory analysis revealed that the characteristics of each mode originates from the different time scales of the reaction, which arise from the different neck relaxation modes controlling the mass drift toward mass symmetry. 
This means that it is possible to discuss the time-dependent functional form of the neck parameter $\epsilon$ during the QF process in the framework of the Langevin equation model. Note that the relaxation of the neck parameter is sensitive to the structure of the QF mass fragment distribution. 
The scission configurations of two modes are both elongated much. However, one of the two modes corresponds to a more elongated scission configuration. The characteristics of two QF modes may lead to useful knowledge in the experimental QF analyses. We hope that the difference of TKE values of QF component in the collision system forming $Z\ge$ 114 targeting actinide nuclei at the medium angular momentum (impact parameter) regions can be confirmed experimentally. 
\\

\begin{acknowledgments}
The Langevin calculation was performed using the cluster computer system (Kindai-VOSTOK) under the supported of JSPS KAKENHI Grant Number 20K04003 and Research funds for External Fund Introduction 2021 provided by Kindai University.
\end{acknowledgments}
%
%


\end{document}